\documentclass[a4paper,10pt]{article}
\usepackage{amsmath}
\usepackage{graphicx}
\usepackage{epsfig}

\title{Quantum critical point in the spin glass-antiferromagnetism
competition in Kondo-lattice systems}

\author{S.G. Magalhaes\footnote{ggarcia@ccne.ufsm.br},
F.M. Zimmer
\\
{\it Departamento de F\'{\i}sica, Universidade Federal de Santa Maria,}
\\ 
{\it 97105-900 Santa Maria, RS, Brazil}
\\
\\
B. Coqblin
\\
{\it Laboratoire de Physique des Solides, Universit\'e Paris-Sud, }
\\{\it b\^atiment 510, 91405 Orsay, France}}

\begin{document}
\date{}
\maketitle


\begin{abstract}

A theory is proposed to describe the competition among
antiferromagnetism (AF), spin glass (SG) and Kondo effect. The model
describes two Kondo sublattices with an intrasite Kondo
interaction strength $J_{K}$ and an interlattice quantum Ising
interaction in the presence of a transverse field $\Gamma$. The
interlattice coupling is a random Gaussian distributed variable
(with average $-2J_0/N$ and variance $32 J^{2}/N$) while the
$\Gamma$ field is introduced as a quantum mechanism to produce
spin flipping. The path integral formalism is used to study this
fermionic problem where the spin operators are represented by
bilinear combinations of Grassmann fields. The disorder is treated
within the framework of the replica trick. The free
energy and the order parameters of the problem are obtained by
using the static ansatz and by choosing both $J_0/J$ and
$\Gamma/J \approx (J_k/J)^2$ to allow, as previously,  a
better comparison with the experimental findings.
  The results indicate the presence of a SG solution at low  $J_K/J$
and for temperature $T<T_{f}$ ($T_{f}$ is the freezing
temperature). When $J_K/J$ is increased, a mixed phase AF+SG appears, then an AF solution and finally  a Kondo state is obtained for high values of $J_{K}/J$.
Moreover, the behaviors of the freezing and Neel
temperatures are also affected by the relationship between $J_{K}$
and the transverse field $\Gamma$. The first one presents a slight
decrease while the second one decreases towards a Quantum Critical
Point (QCP). 
The obtained phase diagram has the same
sequence as the experimental one for $Ce_{2}Au_{1-x}Co_{x}Si_{3}$, 
 if  $J_{K}$ is assumed to increase with $x$,
and in addition, it also shows a qualitative agreement
concerning the behavior of the freezing and the Neel temperatures.

\end{abstract}


\section{Introduction}
\label{Introduction}
The competition between RKKY interaction and Kondo effect is
recognized as the fundamental mechanism to explain several
properties of $Ce$ and $U$ compounds. Nevertheless, there are also
strong experimental evidences that the presence of disorder in
those compounds can be source of quite novel effects
\cite{Bauer,Booth,Maksimov,BenLi,Maple}. For instance, deviations
of the Fermi-Liquid (FL) behavior are observed in the
magnetic, thermodynamic and transport properties. Recently,
theories have been proposed to explain how disorder can produce
such deviations \cite{Miranda,Castro-Neto}. However, it is still
not clear whether disorder itself can be the origin of such
deviations. Other prospective has adopted the proximity to a QCP
\cite{Continentino} as a possible origin of the Non-Fermi Liquid
(NFL) behavior. Besides this complicate scenario with NFL
behavior, some of these physical systems present a highly
non-trivial manifestation of disorder which is frustration. As
consequence, there are many phase diagrams with
ferromagnetism (FM), antiferromagnetism (AF), spin
glass (SG) and a Kondo state with the complete screening of
the localized spins. This raises the question of whether it is
possible to build up theories able to account for
those phase diagrams.

The experimental situation of Ce and U alloys which present
together a spin glass phase, an antiferromagnetic one and finally
a Kondo one is complicate, especially when the phase changes occur
with changing the matrix composition of the alloy. Moreover, there
is certainly an important local effect in the spin glass, as
recently evidenced by local measurements in $CeCu_{1-x}Ni_{x}$ alloys \cite{Marcano}.
In particular, there is a change from AF to
Kondo and then to SG with increasing $x$ in $UCu_{5-x}Pd_{x}$
\cite{Vollmer} and a different situation in
$Ce_{2}Au_{1-x}Co_{x}Si_{3}$ alloys \cite{Majumdar}. More
precisely, in the case of $Ce_{2}Au_{1-x}Co_{x}Si_{3}$ alloys, the
glassy behaviour is not favored when the chemical induced disorder
is initially increased. In the experimental phase diagram, for
$0<x<0.45$ and low temperature, a spin glass-like state appears.
When the $Co$ doping is increasead, the system experiments a
transition to an AF phase. In the interval $0.45<x<0.9$, the Neel
temperature $T_{N}$ decreases until reaching a QCP at $x=0.9$ with
no trace of NFL behavior. For $x>0.9$, the magnetic moments are
supressed due to the Kondo effect.

Recently, a theoretical approach \cite{Magal1} has attempted to
describe the interplay between AF and SG in a Kondo lattice using
the same framework previously introduced to study the  SG in the
Kondo lattice \cite{Alba}. The proposed model has been a
two-Kondo sublattice with an intrasite exchange interation with
strength $J_{K}$ and a random Gaussian $J_{ij}$ intersite
interaction only between distinct sublattices. Two quite important
points are introduced in this approach. The first one is that
there is no hopping of conduction electrons between distinct
sublattices. Therefore, the antiferromagnetic solutions in this
model are entirely associated to the coupling among localized
magnetic moments. The second one takes the relationship
$J_{0}/J\propto (J_{K}/J)^{2}$, because the two exchange terms
cannot be considered as completely independent from each other
\cite{Iglesias}. As a consequence, when the strength of the
intrasite exchange interaction $J_{K}$ increases (in units of
$J$), the degree of frustration $J/J_{0}$ decreases. This last
parameter controls the emergence of SG or AF solutions in the
problem. The results have been shown in a phase diagram $T/J$ ($T$
is the temperature) {\it versus} $J_{K}/J$. For small $J_{K}/J$
and low temperature, there is a SG phase. When $J_{K}/J$ is
increased, an AF solution has been found. Finally, for
$J_{K}/J\geq 15$, the Kondo state becomes dominant. Thus, we
can conclude from the previous calculation \cite{Magal1} that the
sequence of phases SG, AF and Kondo mimics qualitatively the
experimental phase diagram of $Ce_{2}Au_{1-x}Co_{x}Si_{3}$ if
$J_{K}/J$  is associated with the content of $Co$.

But, we can say that the previous agreement was only qualitative
to provide the correct sequence of phases for the alloys
$Ce_{2}Au_{1-x}Co_{x}Si_{3}$ \cite{Majumdar}, because in the
previous approach 
there is no mechanism able to produce a QCP. As a consequence, this approach
cannot provide the proper behavior of the Neel temperature which
should decrease towards a QCP. Thus, the purpose of the present
paper is to obtain specifically a quantitative description of the
phase diagram of $Ce_{2}Au_{1-x}Co_{x}Si_{3}$ alloys and in
particular of the QCP, by adding a quantum mechanism to produce
spin flipping, given by an additional transverse field $\Gamma$
\cite{AlbaCoqblin}.

We take here the replica symmetry mean-field approximation and we use the static
approximation. Indeed, such an approximation is subject to
criticism, but it was shown in many previous works that it yields
a good description of the competition between the Kondo effect and
magnetism in the Kondo lattice models. Moreover, the quantum Ising
spin glass in a transverse field has the same critical behavior as
the $M$-component quantum rotor which is exactly solvable in the
$M\rightarrow \infty$ limit \cite{Sachdev}. In this model, the
line transition is given by the singularity of the zero-frequency
that is equivalent to the static approximation. The static
approximation is, therefore, relatively well justified for the
study of phase boundaries, which is really the purpose of our
paper in the specific case of the SG-AF-Kondo competition in
alloys  such as $Ce_{2}Au_{1-x}Co_{x}Si_{3}$.

Then, we will assume a relationship $\Gamma\propto J_{K}^{2}$;
this choice, which has been already used in references
\cite{AlbaCoqblin,Iglesias}, 
is taken to account for the fact that the
intrasite exchange interaction is able to produce both the Kondo
effect and the RKKY interaction. 
The transverse field plays a role
similar to the spin flipping part of the Heisenberg model. One of
the main achievements of this approach has been to show the
presence of a QCP in the competition between SG and the Kondo
state when $\Gamma$ is enhanced and to improve, therefore,
considerably the description of experimental results.

In this paper, we study the SG-AF competition in the disordered
Kondo lattice, described by the hopping of the conduction
electrons only inside each sublattice, an intersite exchange
interaction, given by an Ising-like term only between different
sublattices and a transverse field $\Gamma$ applied in the
$x$-direction, in order to obtain a mechanism able to produce a
QCP. We use also the static approximation and the relationships
$J_{0}\propto J_{K}^{2}$ and $\Gamma\propto J_{K}^{2}$ already
introduced in references \cite{Magal1,AlbaCoqblin}. Therefore, the
initial many parameters problem is reduced to a one parameter
which allows a better comparison with experimental results.

The comparison of our theoretical results to experimental data is delicate for two reasons. First, as we have just discussed, we take a $J_{K}^{2}$-dependence of both $J_{0}$ and $\Gamma$, which is clearly a crude approximation but which has already given interesting physical results in reference \cite{Iglesias}. Then, the second question concerns the fact that $J_{K}$ is assumed to increase with the concentration of cobalt in $Ce_{2}Au_{1-x}Co_{x}Si_{3}$ alloys, as previously done with the concentration of Nickel  in $CeCu_{1-x}Ni_{x}$ alloys. Indeed, the role of the so-called "chemical pressure"  is less clear than that of the regular pressure, but obviously the spin glass phase originates in such alloys only from the disorder of the matrix and we are obliged to make an assumption of the relationship between $J_{K}$ and the relative host concentration x, which was previously successful in the case of $CeCu_{1-x}Ni_{x}$ alloys  reference \cite{AlbaCoqblin}. This question is in fact not easy to answer theoretically.

In section 2, we introduce the model and the relevant order
parameters for the problem. The saddle point free energy in terms
of the order parameters is obtained so that it allows to obtain a
phase diagram giving temperature $T/J$ {\it versus} $J_{K}/J$. The
results and the final discussion are presented in the last
section.
\section{General Formulation}
\label{GF}
We have considered in this work a model given by two Kondo
sublattices $A$ and $B$ with a random coupling $J_{ij}$ only
between localized spins in distinct sublattices \cite{Magal1,Korenblit}.
There is also a transverse field
$\Gamma$ coupled with the localized spins in both sublattices \cite{AlbaCoqblin}.
As mentioned in section 1, the hopping of
conduction electrons between two different sublattices is not
allowed for simplicity. Therefore, the Hamiltonian is:
\begin{align}
 H-\mu N=\sum_{p=A,B}[\sum_{i,j}\sum_{\sigma=\uparrow \downarrow}
 t_{i j}\hat{d}_{i,p,\sigma}^{\dagger}\hat{d}_{j,p,\sigma} +
\sum_{i}\varepsilon_{0, p}^{f}\hat{n}_{i,p}^{f}~~
\nonumber\\+
J_{K}(\sum_{i}\hat{S}_{i,p}^{+}\hat{s}_{i,p}^{-}+\hat{S}_{i,p}^{-}\hat{s}_{i,p}^{+})]+
\sum_{i, j} J_{i j}\hat{S}_{i,A}^{z}\hat{S}_{j,B}^{z}+2\Gamma\sum_{i} (\hat{S}_{i,A}^{x}+\hat{S}_{i,B}^{x})~~
\label{e1}
\end{align}
\noindent where $i$ and $j$ sums are over $N$ sites of each
sublattice. The intersite coupling $J_{i j}$ is a random variable
following a gaussian distribution with average $-2J_0/N$ and
variance $32 J^{2}/N$. The spin operators present in Eq.\
(\ref{e1}) are defined as in Ref. \cite{Magal1}:
$\displaystyle \hat{s}_{i,p}^{+}=\hat{d}_{i,p,\uparrow}^{\dagger}
\hat{d}_{i,p,\downarrow}=(\hat{s}_{i,p}^{-})^{\dagger}
$, $\displaystyle \hat{S}_{i,p}^{+}=\hat{f}_{i,p,\uparrow}^{\dagger}
\hat{f}_{i,p,\downarrow }=(\hat{S}_{i,p}^{-})^{\dagger}
$,
$\hat{S}_{i,p}^{z}= \frac{1}{2}
\left[\hat{f}_{i,p,\uparrow}^{\dagger}
\hat{f}_{i,p,\uparrow }\right.$ $-\hat{f}_{i,p,\downarrow}^{\dagger}$
$\left.\hat{f}_{i,p,\downarrow }\right]$
and
$\hat{S}_{i,p}^{x}= \frac{1}{2}
\left[\hat{f}_{i,p,\uparrow}^{\dagger}
\hat{f}_{i,p,\downarrow}\right.$ $+\hat{f}_{i,p,\downarrow}^{\dagger}$
$\left.\hat{f}_{i,p,\uparrow }\right]$ where $\hat{d}_{i,p, \sigma}^{\dagger}$, $\hat{d}_{i,p,\sigma}$ $(\hat{f}
_{i,p, \sigma}^{\dagger},\hat{f}_{i,p, \sigma})$
are the creation and destruction operators for conduction (localized) fermions.

The partition function is given in the 
Lagrangian 
path integral
formalism in terms of Grassman variables \cite{Magal1,Alba}
$\psi_{i,p,\sigma}(\tau)$ for the localized fermions and
$\varphi_{i,p,\sigma}(\tau)$ for the
conduction
ones. The Kondo
interaction is treated following Ref. \cite{Magal1}. 
The Kondo state is described by complex order parameters
$\lambda_{p,\sigma}=\frac{1}{N}\sum_{i,\sigma}<\varphi^{*}_{i,p,\sigma}\psi_{i,p,\sigma}>$
which are introduced in the partition function using the
integral representation of delta function. Therefore, the
partition function when $\lambda_{p\sigma}\approx \lambda_{p}$
becomes:
\begin{equation}
Z/Z_{0}^{d}=\exp\{-2N\beta J_k(|\lambda_A|^2+|\lambda_B|^2)\}\int\!\!\prod_{p=A,B} D(\psi_{p}^{*}\psi_{p})
\exp\left[ A_{eff}^{(stat)} \right]
\label{e121}
\end{equation}
where $Z_{0}^{d}$ is the partition function of the free conduction electrons,
\begin{equation}
A_{eff}^{(stat)}=\sum_{i,j}\sum_{\omega}
\underline{\Theta}_{i}^{\dagger}(\omega)
\left[\underline{\underline{g}}_{i j}(\omega)\right]^{-1}
\underline{\Theta}_{j}(\omega)+\beta \sum_{ij} J_{ij} S^{z}_{i,A} S^{z}_{j,B},
\label{e14100}
\end{equation}
\begin{equation}
\underline{\Theta}_{i}^{\dagger}(\omega)=
\left[\begin{tabular}{cc}
$\psi_{i,A,\uparrow}^{*}(\omega)$ $\psi_{i,A,\downarrow}^{*}(\omega)$ $\psi_{i,B,\uparrow}^{*}(\omega)$ $\psi_{i,B,\downarrow}^{*}(\omega)$
\end{tabular}\right]
\end{equation}
and $\left[\underline{\underline{g}}_{i j}(\omega)\right]^{-1}$ is a $4 \times 4$
matrix given by
\begin{equation}
\left[\underline{\underline{g}}_{i j}(\omega)\right]^{-1} =
\left[\begin{tabular}{cc}
$\underline{\underline{F}}_{A,ij}(\omega)$  & $\underline{\underline{0}}$ \\
$\underline{\underline{0}}$ & $\underline{\underline{F}}_{B,ij}(\omega)$
\end{tabular}\right]
\label{e150}
\end{equation}
with
\begin{equation}
\underline{\underline{F}}_{p,ij}(\omega)=
\left[\begin{tabular}{cc}
$(i\omega -\beta\varepsilon_{0,p}^{f})$ & $\beta\Gamma$ \\
$\beta\Gamma$ & $(i\omega -\beta\varepsilon_{0,p}^{f})$
\end{tabular}\right]\delta_{i,j}-\sum_{k} \frac{\beta^2 J_{K}^{2}|\lambda_{p}|^{2} e^{i\vec{k}(\vec{R}_{i}-\vec{R}_{j})}}
{(i\omega-\beta\varepsilon_0)-\beta\epsilon_k}\underline{\underline{I}}.
\label{ef}
\end{equation}
The notations $\underline{\underline{I}}$ and $\underline{\underline{0}}$
represent the identity and the null $2\times 2$ matrices, respectively.

The free energy is given by the replica method. The procedures are
quite close to reference \cite{Magal1}  and in particular to
its appendix, but with now the presence of the transverse field
$\Gamma$ which modifies the matrix
$\left[\underline{\underline{g}}_{i j}(\omega)\right]^{-1}$.
Therefore, the averaged free energy within the replica symmetric
theory can be obtained as
\begin{align}
\beta F&=
\beta J_K(|\lambda_A|^2+|\lambda_B|^2)- \beta^2\frac{J^2}{2} (q_A q_B-
\tilde{q}_A \tilde{q}_B)
-\beta \frac{J_0}{2} m_A m_B-\lim_{n\rightarrow 0}\frac{1}{2Nn}
\nonumber\\
\times
\ln&
\int_{-\infty}^{\infty}\prod_{i=1}^{N}D\xi_{i,A}
D\xi_{i,B}
\prod_{\alpha}\int_{-\infty}^{\infty}D z_{i,A}^{\alpha}  D z_{j,B}^{\alpha}
\exp\left[\sum_{w,\sigma}\ln \det \left[\underline{\underline{G}}_{ij}(\omega|h^{\alpha}_{i,p})\right]^{-1}\right]
\label{e25}
\end{align}
\noindent where the matrix $\left[\underline{\underline{G}}_{ij}(\omega|h^{\alpha}_{i,p})\right]^{-1}$
in Eq.\ (\ref{e25}) is the same matrix given in Eq. (\ref{e150}), except that the matrix
$\underline{\underline{F}}_{i j,p}(\omega)$  (see Eq. (\ref{ef})) is replaced by
$\underline{\underline{F}}_{i j,p}(\omega)+\sigma\beta J h_{i,p}^{\alpha}\delta_{ij}
\underline{\underline{I}}$ ($\sigma=+ (\uparrow), - (\downarrow$)).
In Eq. (\ref{e25}) $D y= dy \exp(-y^2/2)/$ $\sqrt{2\pi}$ $(y=\xi, z)$.
The internal field $h_{i,p}^{\alpha}$ applied in the sublattice $p=A,B$ is defined as
\begin{equation}
h_{i,p}^{\alpha}= J\sqrt{2q_{p'}}\xi_{i,p}+J\sqrt{2\overline{\chi}_{p'}}z^{\alpha}_{i,p}- J_{0}m_{p'}.
\label{internalfield}
\end{equation}
In Eq. (\ref{internalfield}),  $q_{p'}$,
$\chi_{p'}=\beta\overline{\chi}_{p'}$ and $m_{p'}$ are the SG
order parameter, the static susceptibility and the magnetization,
respectively. It should be noticed that the internal field applied
in sublattice $p$ depends on the behavior of $q_{p'}$, $\chi_{p'}$
and $m_{p'}$ with $p'\neq p$. That is a direct consequence
of the choice made in the model where only interlattice
frustration has been considered in the present work.

Finally, to solve Eq. (\ref{e25}), the matrix
$\left[\underline{\underline{G}}_{ij}(\omega|h^{\alpha}_{i,p})\right]^{-1}$
is substituted by
$\left[\underline{\underline{G}}_{\mu\nu}(\omega|\right.$ $\left.h^{\alpha}_{i,p})\right]^{-1}$.
In fact, this approximation \cite{Magal1,Alba,AlbaCoqblin}
represents a Kondo sublattice $p$ in the presence of a constant
magnetic field $h^{\alpha}_{i,p}$ where the effects of the Kondo
sublattice $p'$ are placed. Therefore, in Eq. (\ref{e25}), we have:
\begin{equation}
\ln \det \left[\underline{\underline{G}}_{ij}(\omega|h^{\alpha}_{i,p})\right]^{-1}=
\frac{1}{N}\sum_{j}\ln \det \left[\underline{\underline{G}}_{\mu\nu}(\omega|h^{\alpha}_{i,p})\right]^{-1}.
\label{decoupling}
\end{equation}
The sum over the Matsubara's frequencies in Eq. (\ref{e25}) can be
now performed as usual \cite{Magal1,Alba,AlbaCoqblin}.
The proposed decoupling also allows us to use
Fourier transformation which can be evaluated by assuming
a constant density of states for the conduction electron band $\rho(\epsilon)= 1/2D$ for
$-D<\epsilon<D$. Thereby, the free energy is obtained as:
\begin{align}
\beta F&= \beta J_K(|\lambda_A|^2+|\lambda_B|^2) + \frac{\beta^2 J^2}{2} \bar{\chi}_A \bar{\chi}_B
+\frac{\beta^2 J^2}{2} (\bar{\chi}_{A}q_B+\bar{\chi}_B q_A)
\nonumber\\
&-\frac{\beta J_0}{2} m_A m_B-
\frac{1}{2} \int_{-\infty}^{\infty}\!\!D\xi_{i_A}\!
\int_{-\infty}^{\infty}\!\!D\xi_{j_B}
\ln\left[\prod_{p=A,B}
\int_{-\infty}^{\infty}D z_{p}e^{E( H_{p})}\right]
\label{e28}
\end{align}
where
\begin{equation}
E(H_{p})=\frac{1}{\beta D}\int^{+\beta D}_{-\beta D} dx\ln\{\cosh(\frac{x+\beta H_{p}}{2})+
\cosh(\sqrt{\Delta}) \}
\label{E(H_{p})}
\end{equation}
with
\begin{equation}
\Delta=\frac{1}{4}(x-\beta H_{p})^{2}+(\beta J_{K} \lambda_{p})^{2}
\label{D}
\end{equation}
and $H_{p}=\sqrt{\Gamma^{2}+h_{p}^{2}}$, with the internal
field $h_{p}$ given by Eq. (\ref{internalfield}) (without here the
site and replica indices). The saddle point order parameters
follow directly from equations (\ref{e28})-(\ref{D}).

The limit of stability for the order parameters with replica symmetry is given when 
the Almeida-Thouless eigenvalue $\lambda_{AT}$ becomes negative \cite{AT, Magal2}:   
\begin{equation}
\lambda_{AT}=1-2(\beta J)^4\prod_{p=A,B} \int_{-\infty}^{\infty} D\xi_p\left[\frac{I_p(\xi_p)}{(\int_{-\infty}^{\infty}Dz_p {\mbox e}^{E(H_p)})^2}\right]^2
\label{AT}
\end{equation}
where
\begin{equation}
I_p(\xi_p)=\int_{-\infty}^{\infty}Dz_p {\mbox e}^{E(H_p)}\int_{-\infty}^{\infty}Dz_p\frac{\partial}{\partial h_p}[{\mbox e} ^{E(H_p)}
\frac{\partial E(H_p)}{\partial h_p}]-[\int_{-\infty}^{\infty}Dz_p{\mbox e} ^{E(H_p)}
\frac{\partial E(H_p)}{\partial h_p}]^2.
\end{equation}

\section{Results and Conclusions}
\label{RC}

The central issue of the present work is to account for the
general features of the experimental phase diagram of $Ce$ alloys
such as $Ce_{2}Au_{1-x}Co_{x}Si_{3}$ \cite{Majumdar}  which
present with increasing $Co$ concentration the successive sequence
of the SG phase and then the AF phase with the Neel temperature
$T_{N}$ decreasing towards a QCP. In principle, the solutions for
the saddle point order parameters $q_{p}$, $\overline{\chi}_{p}$,
$m_{p}$ and $\lambda_{p}$ ($p=A$, $B$) should be found in a
parameter space with axes given by $J_{0}/J$, $J_{K}/J$ and
$\Gamma/J$ ($J$ and $D$  are kept constant). 

\begin{figure}
\begin{center}
    \includegraphics[angle=270,width=13cm]{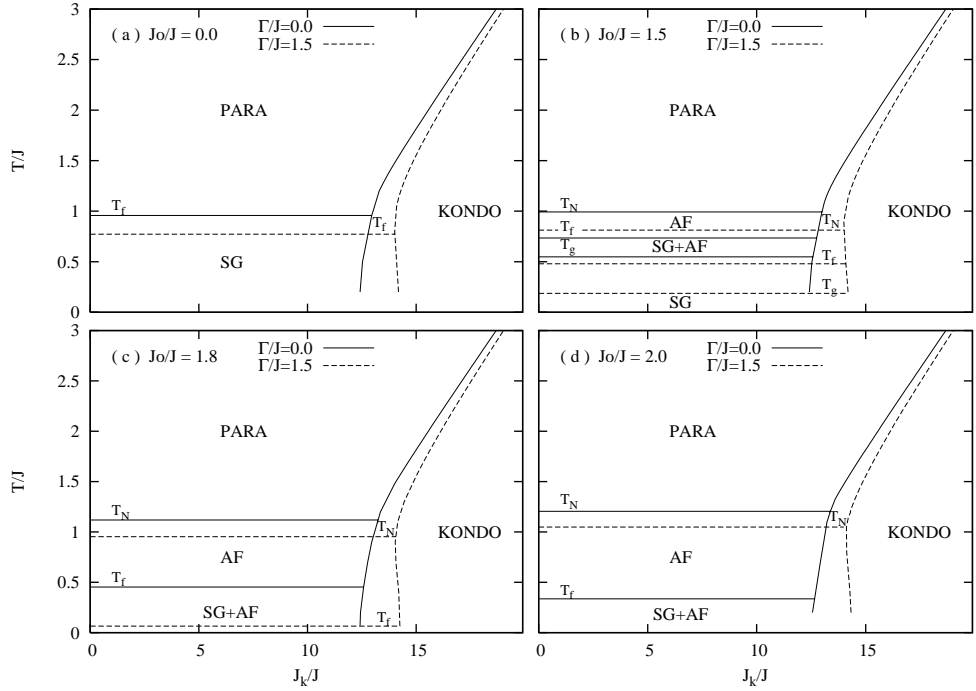}
\caption{ Theoretical phase diagram $T/J$ {\it versus} $J_k/J$ for several values of $J_0/J$:
(a) $J_0/J=0$, (b) $J_0/J=1.5$, (c) $J_0/J=1.8$ and (d) $J_0/J=2.0$.
The full lines correspond to the results obtained for $\Gamma/J=0.0$, while the dashed ones are for $\Gamma/J=1.5$. 
} \label{fig2}
\end{center}
\end{figure}

Therefore, in order to identify the role of each parameter in the problem,  
the solutions for $q_{p}$, $\overline{\chi}_{p}$,
$m_{p}$ and $\lambda_{p}$ ($p=A$, $B$)  are shown in figure 1 as a 
phase diagram $T/J$ {\it versus} $J_{k}/J$ for several 
values of $J_{0}/J$ and $\Gamma/J$. 
The SG solution corresponds to $q_{A}=q_{B}\neq
0$ (with $m_{A}=m_{B}=0$) while the AF solution is equivalent of
$m_{A}=-m_{B}\neq 0$ and $q_{A}=q_{B}\neq 0$. 
The Kondo state in the phase diagram
corresponds to the situation where the two $\lambda_{p}$ ($p=A,B$) are
different from zero. 
In figure 1.a, $J_{0}=0$ and $\Gamma=0$ or $\Gamma/J=1.5$, this situation corresponds 
to the studied case in Ref. \cite{AlbaCoqblin} in which 
there is no AF solution. 
For $\Gamma=0$, the phase diagram displays three solutions. 
At high temperature and small $J_{K}/J$, a paramagnetic (PARA) solution is found. 
When the temperature is decreased, there is 
a transition  to the SG phase at $T_{g}$, 
which coincides with the $T_{f}$ temperature 
that locates the Almeida-Thouless line (the temperature where the Almeida-Thouless eigenvalue is zero). 
If $J_{K}/J$ is enhanced, there is a 
transition at 
$J_{K}=J_{K_{1}}(T/J)$ 
to a region where the 
Kondo solution is dominant. 
When $\Gamma/J$ is tuned on, there is a clear effect on  
the transition lines $T_{g}$ and $J_{K_{1}}(T/J)$, 
 the first one is decreased while the second is displaced
 to a larger $J_{K}/J$ value.  

When $J_{0}/J$ is enhanced, as it is shown in figures 1.b - 1.d, an AF solution for the order parameters
appears below the Neel temperature $T_N$ for small $J_{K}/J$. 
For this particular situation, 
the set of temperatures 
$T_N$, $T_{f}$ and $T_{g}$ is 
function of $\Gamma/J$ and $J_0/J$. 
For $\Gamma=0$, the temperature $T_N$
increases with $J_{0}/J$ while the replica symmetric transition temperature $T_{g}$ and the Almeida-Thouless
line $T_{f}$ decrease
with $J_{0}/J$. Nevertheless, similar to Ref. \cite{Magal2}, $T_{f}$ becomes larger than $T_{g}$. 
Therefore, in such circumstances, 
there is the onset of a mixed phase AF+SG at $T_{f}$. 
This mixed phase represents a replica symmetry breaking SG phase with non-zero 
spontaneous sublattice magnetization. Finally, at $T_{g}$ there is a transition to a pure SG phase. For 
increasing $J_{K}/J$,  a Kondo solution at 
$J_{K}>J_{K_{1}}(T/J)$ 
is found again. When the 
transverse field
$\Gamma/J$ is tuned on, 
its effect is basically to decrease simultaneously $T_N$, $T_{f}$ and $T_{g}$ 
and displace the transition line $J_{K_{1}}(T/J)$ as before.

However, 
to compare the results obtained in this work
with the experimental results, it is better to reduce the number
of parameters which are in fact not independent from each other,
because both the Kondo effect and the RKKY can be generated
directly from the intrasite exchange interaction. Thus, we choose
a relationship among the parameters $J_{0}/J$, $J_{K}/J$ and
$\Gamma/J$, similar to that used in Ref. \cite{Magal1} where a
relationship $J_{0}/J\propto (J_{K}/J)^{2}$ has successfully
reproduced the experimental sequence of phases. Nevertheless, in
Ref. \cite{Magal1}, we cannot explain at all the decrease of the
Neel temperature $T_{N}$ towards a QCP. In the present work, the
transverse field $\Gamma$ has been introduced to provide a simple
mechanism able to reproduce the spin flipping part of the
Heisenberg model \cite{AlbaCoqblin}. Therefore, it would be also
natural to assume $\Gamma/J\propto (J_{K}/J)^{2}$ in the present
problem. It should be remarked that this choice introduces an
additional complexity. It has been shown that $\Gamma/J$ and
$J_{0}/J$ enforce different effects in a previously studied
competition between AF and SG. Within a certain range of $J_{0}/J$,
this new choice can favor SG, SG+AF or AF, while $\Gamma/J$ tends to
destroy the three phases leading the transition temperatures to zero
\cite{Zimmer}.

\begin{figure}
\begin{center}
    \includegraphics[angle=0,width=11.5cm]{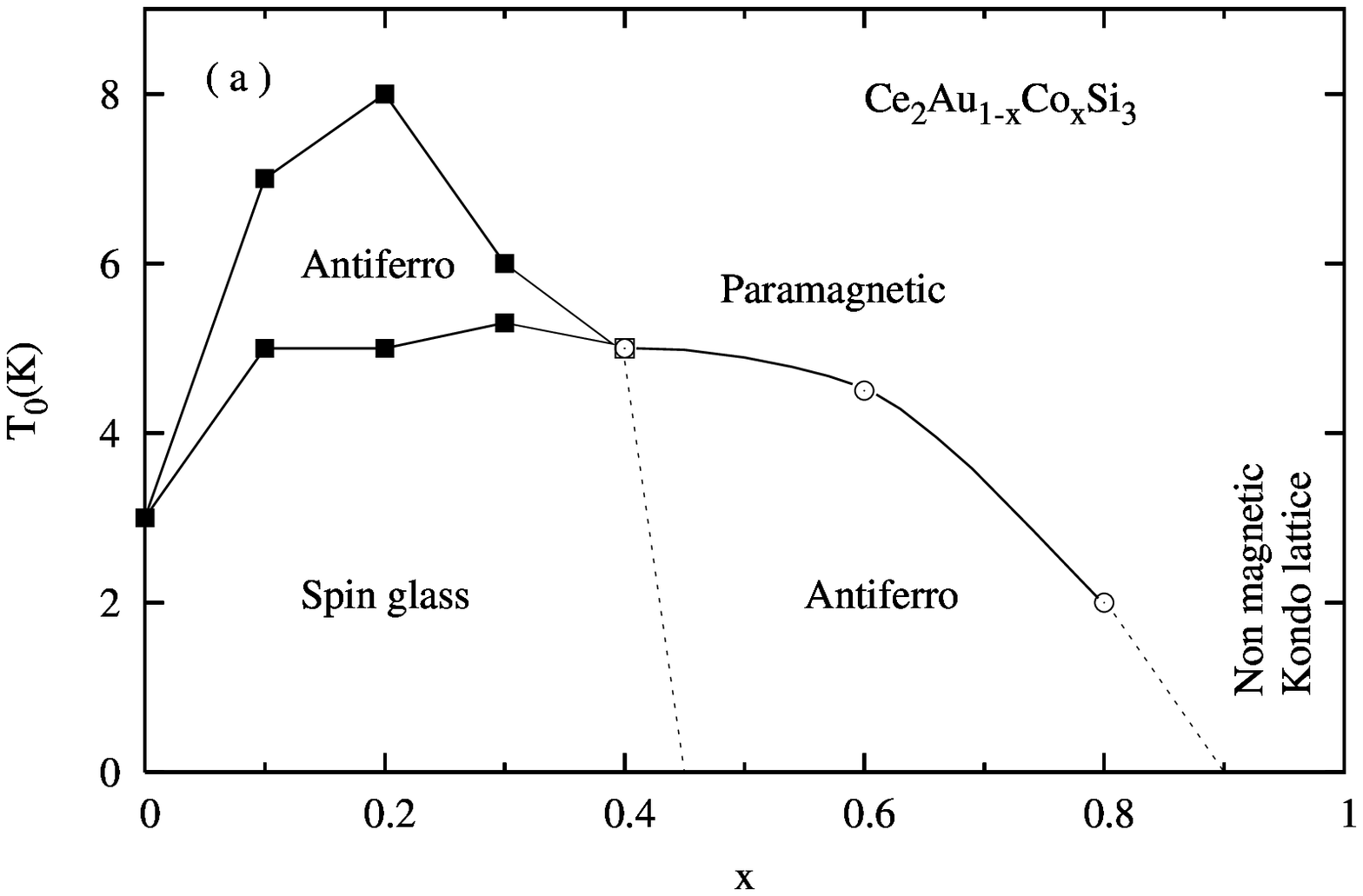}
    \includegraphics[angle=0,width=11.5cm]{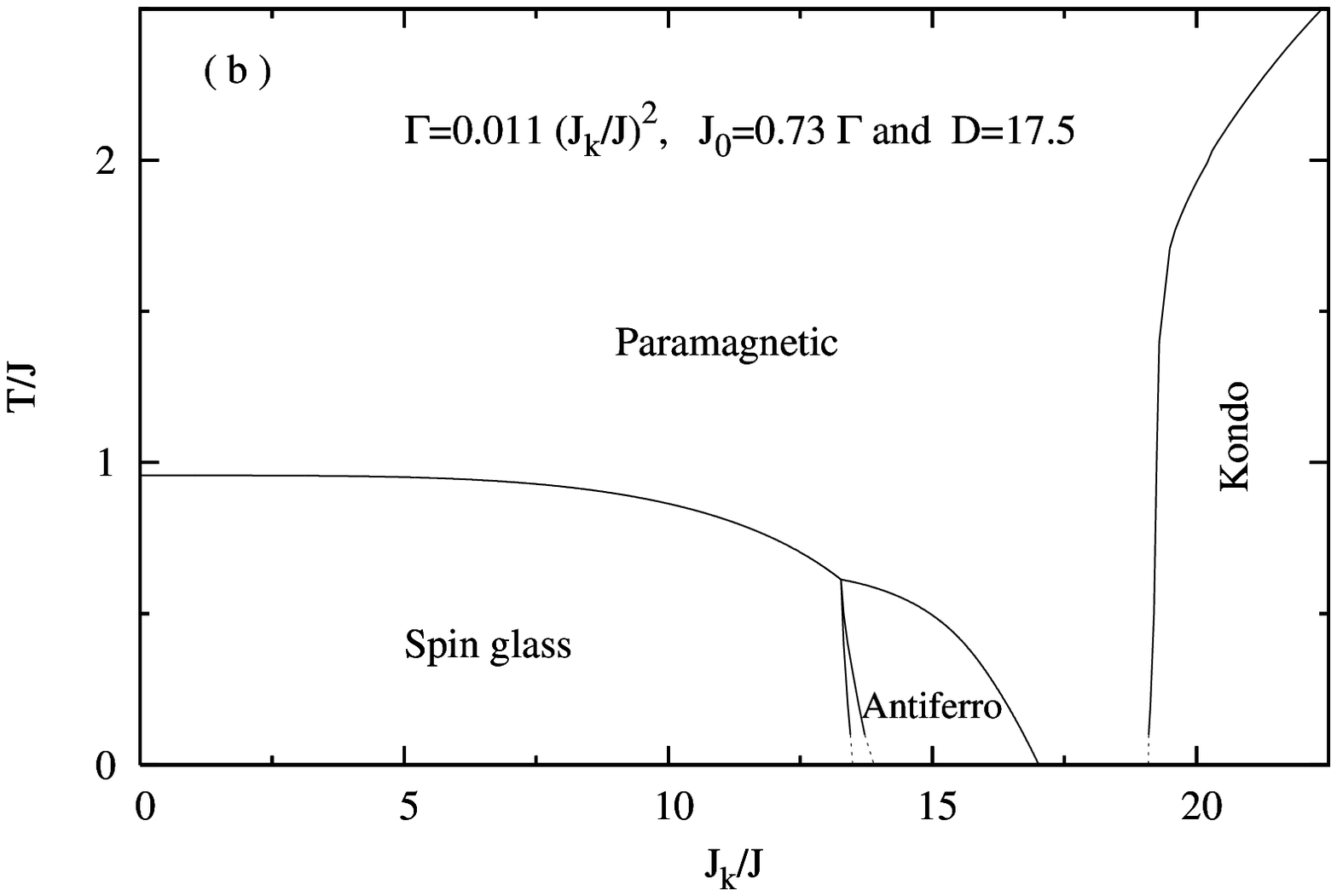}
\end{center}
\caption{Comparison between experimental and theoretical
phase diagrams: the upper figure is an experimental phase diagram
of $Ce_2 Au_{1-x} Co_x Si_3$ alloys \cite{Majumdar}; the lower
figure is a theoretical phase diagram $T/J$ {\it versus} $J_{K}/J$
for the relations $\Gamma= 0.011 (J_{K}/J)^{2}$ and $J_{0}/J=0.73
\Gamma/J$, where the dotted lines are the extrapolations carried
for lower temperature.} \label{fig1}
\end{figure}

The solution of the set $q_{p}$, $\overline{\chi}$, $m_{p}$
and $\lambda_{p}$ with $J_{0}/J\propto \Gamma/J\propto (J_{K}/J)^{2}$ is
shown in figure 2.b giving the phase diagram 
$T/J$ {\it
versus} $J_{K}/J$. 
In particular, the Neel temperature
$T_{N}\equiv T_{N}(J_{K})$ can be computed expanding the saddle
point equations in powers of $m_p$. At the second order critical
line $T_{N}$, we can make $q_p=\lambda_{p}=0$, therefore
$\bar{\chi}=\bar{\chi}_{p}=1/(\beta_{c} J_{0})$ and
\begin{equation}
m\left[1-\beta_c J_0\left(\frac{1+\int D\xi \xi^2 \cosh\beta_{c}
\sqrt{\frac{2 J \xi^2}{J_0 \beta_{c}}+\Gamma^2}}{1+
\int D\xi \cosh\beta_{c}\sqrt{\frac{2 J \xi^2}{J_0 \beta_{c}}+\Gamma^2}}\right)\right]=0
\label{mc}~,
\end{equation}
where $\beta_c =1/T_{N}$ and $m=m_A=-m_{B}$. For $T_{N}$ tending
to zero, the critical value of $\Gamma_{c}$ is given by analytical
solution of Eq. (\ref{mc}): $\Gamma_{c}= J_0 + 2 J^2/J_0$.  For
 $0\leq J_{K}/J \leq 13.5$, figure 2.b shows the presence of the SG
solution in which $T_g$  and 
$T_f$ coincide. Then, in a very small interval 
of $J_{k}/J$, $T_f \neq T_{g}$.
Therefore, the mixed phase 
AF+SG  appears 
after the SG phase.       
Finally, in the interval $13.9\leq J_{K}/J\leq 17.0$, only the AF
solution is present. When $J_{K}/J$ increases, the Neel temperature decreases until a QCP.
Above this point onwards, there is a transition line to the Kondo
state characterized by $\lambda_{A}=\lambda_{B}\neq 0$
\cite{AlbaCoqblin}. At high temperature, the solutions of the
order parameters are $q_{p}=0$, $m_{p}=0$ and $\lambda_{p}=0$
($p=A$, $B$).

The situation described in the present replica symmetry mean field theory can be
analysed as follows: for a low $J_{K}/J$ value (corresponding to
high value of $J/J_{0}$), frustration is dominant and the long
range internal field is given by the SG component of $h_{p}$ (see
Eq. (\ref{internalfield})). When $J_{K}/J$ is increased, the
degree of frustration $J/J_{0}$ is  decreased and, therefore, the
localized spin of the sublattice $A$ ($B$) starts to couple with
an internal field which depends on the negative magnetization of
the sublattice $B$ ($A$) in the AF ordering \cite{Magal1}.
However, this process also implies the increase of the transverse
field $\Gamma$ and consequently the freezing temperature ($T_{g}$)
is slightly decreasing. On the other hand, the Neel temperature
($T_{N}$) is deeply affected and decreases towards a QCP. Finally,
for enough high $J_{K}/J$, there is complete dominance of the
Kondo effect \cite{Magal1,Alba,AlbaCoqblin}. Finally, 
 as we have already explained in the introduction, the
comparison between theory and experiment is 
more delicated 
in the case
of disordered alloys where the different phases occur as a function of the relative matrix concentration than in the classical case where increasing pressure
makes $J_{K}$ increase. But here we consider that $J_{K}$
increases with increasing $Co$ concentration , as we have successfully assumed that $J_{K}$
increases with increasing $Ni$ concentration in the case of $CeCu_{1-x}Ni_{x}$ alloys reference \cite{AlbaCoqblin}.
Under this
assumption, the theoretical results given in figure 2.b agree
quite well with the experimental phase diagram of
$Ce_{2}Au_{1-x}Co_{x}Si_{3}$ alloys \cite{Majumdar} (see
figure  2.a) and in particular we obtain the same order for the
sequence of phases and, mainly, the  correct behavior of the Neel
temperature.

To conclude, we have used a two-sublattice model with two exchange
interactions, an intrasite exchange and a random intersite between
localized Ising spins in the presence of transverse field. The
main goal of the present work has been to reproduce some
fundamental aspects of the phase boundary contained in the
experimental phase diagram of the $Ce_{2}Au_{1-x}Co_{x}Si_{3}$
alloys \cite{Majumdar}. Finally, the theoretical results
shown in figure 2.b account quite well for the most important part
of the experimental phase diagram of $Ce_{2}Au_{1-x}Co_{x}Si_{3}$
alloys \cite{Majumdar} with the cobalt concentration (figure {2.a})
and the agreement between the two figures shows a clear
improvement performed by the present model with respect to
previous ones. In fact, important questions are not really solved,
such as for example the different possible Kondo-AF-SG sequences
found in different disordered alloys, the relationship between the
parameters of the model and the matrix concentration which is
certainly less clear than the pressure dependence and finally the
precise local nature of the spin glass phase. Further work will
be, therefore, necessary to improve the theoretical description
and to obtain new phase diagrams of $Ce$ or $U$ alloys or
compounds with either the matrix concentration or the external
pressure.




\begin{thebibliography}{99}

\bibitem{Bauer} Bauer E D, Booth C H,  Kwei G H,  Chau R
                and Maple M B, Phys. Rev. B {\bf 65}, 245114 (2002).
\bibitem{Booth} Booth C H, Scheidt E W, Killer U, Weber A and Kehrein S,
          Phys. Rev. B {\bf 66}, 140402 (2002).

\bibitem{Maksimov} Maksimov I, Litterst F J, Rechemberg H, Melo M A C,
                   Feyerherm R, Hendrikx R W A, Goertenmulder T J, Mydosh J A
                   and  S\"ullow S, Phys. Rev. B {\bf 67}, 104405 (2003).
\bibitem{BenLi} Young Ben-Li, MacLaughlin D E, Rose M S, Ishida K, Bernal O O,
              Lukefhar H G, Heuser K, Stewart G R, Butch N P, Ho P C
                 and Maple M B,  Phys. Rev. B {\bf 70}, 024401 (2004).

\bibitem{Maple} Zapf V S, Dickey R P, Freeman E J, Suivent C and Maple M B,
               Phys. Rev. B {\bf 61}, 024437 (2001).

\bibitem{Miranda} Miranda E and Dobrosavljevic V, Phys. Rev. Lett. {\bf 86}, 264 (2001).

\bibitem{Castro-Neto} Castro Neto A H and Jones B A, Phys. Rev. B {\bf 62}, 14975 (2000).
	  
\bibitem{Continentino} Coleman P, Physica B {\bf 259-261}, 353 (1999).

\bibitem{Marcano} Marcano N, Espeso J I, Gomez-Sal J C, Fernandez J R,
                    Herrero-Albillos J and Bartolome F,  Phys. Rev. B 
		     {\bf 71}, 134401 (2005).
\bibitem{Vollmer} Vollmer R, Pietrus T, Lohneysen H V, Chau R and Maple M B,
                    Phys. Rev. B {\bf 61}, 1218 (2000).

\bibitem{Majumdar} Majundar S, Sampathkumaran E V, Berger St, Della Mea M,
               Michor H, Bauer E, Brando M, Hemberger J and Loidl A,
             Solid State Commun. {\bf 121}, 665 (2002).

\bibitem{Magal1} Magalhaes S G, Schmidt A A, Zimmer F M, Theumann A and Coqblin B,
                 Eur. Phys. J. B {\bf 34}, 447 (2003).
\bibitem{Alba} Theumann Alba, Coqblin B, Magalhaes S G and Schmidt A A, Phys. Rev. B
                 {\bf 63}, 054409 (2001).

\bibitem{AT} Almeida J R L, Thouless D J, J. Phys. A: Math. Gen. {\bf 11}, 983 (1978).

\bibitem{Magal2} Magalhaes S G, Schmidt A A, F M, Theumann A and Coqblin B,
                 {Eur. Phys. J.} B {\bf 30}, 419 (2002).

\bibitem{Iglesias} Iglesias J R, Lacroix C and Coqblin B, 
                   { Phys. Rev.} B {\bf 56}, 11820 (1997).
		   
\bibitem{AlbaCoqblin} Theumann Alba and Coqblin B,{Phys. Rev.} B {\bf 69}, 214418 (2004).

\bibitem{Sachdev} Ye J, Sachdev S and Read N, {Phys. Rev. Lett.} {\bf 70}, 4011 (1993).

\bibitem{Korenblit} Korenblit I Ya and Shender E F, {Sov. Phys. JETP}
                   {\bf 62}, 1030 (1985).

\bibitem{Zimmer} Theumann Alba, Schmidt A A and Magalhaes S G, {
                Physica} A {\bf 311}, 498 (2002).
\\ 
Zimmer F M and S. G. Magalhaes, Physica A {\bf 359}, 380 (2006).
\end{thebibliography}
\end{document}